\documentclass{ws-ijtaf}

\usepackage{amssymb,eurosym,cite,colortbl,subfigure,multirow}

\begin{document}

\markboth{D. J. Fenn et al.}
{Triangular Arbitrage in the Foreign Exchange Market}


\title{THE MIRAGE OF TRIANGULAR ARBITRAGE\\
IN THE SPOT FOREIGN EXCHANGE MARKET}

\author{DANIEL J. FENN and SAM D. HOWISON}
\address{Oxford Centre for Industrial and Applied Mathematics\\
Mathematical Institute, University of Oxford
\\Oxford OX1 3LB, U.K.\\
fenn@maths.ox.ac.uk\\ howison@maths.ox.ac.uk}

\author{MARK MCDONALD and STACY WILLIAMS}
\address{FX Research and Trading Group, HSBC Bank\\
8 Canada Square, London E14 5HQ, U.K.\\
mark.mcdonald@hsbcib.com\\
stacy.williams@hsbcgroup.com}

\author{NEIL F. JOHNSON}
\address{Physics Department, University of Miami\\
Coral Gables, Florida 33146, U.S.A.\\
njohnson@physics.miami.edu}

\maketitle


\begin{abstract}
We investigate triangular arbitrage within the spot foreign exchange market using high-frequency executable prices. We show that triangular arbitrage opportunities do exist, but that most have short durations and small magnitudes. We find intra-day variations in the number and length of arbitrage opportunities, with larger numbers of opportunities with shorter mean durations occurring during more liquid hours. We demonstrate further that the number of arbitrage opportunities has decreased in recent years, implying a corresponding increase in pricing efficiency. Using trading simulations, we show that a trader would need to beat other market participants to an unfeasibly large proportion of arbitrage prices to profit from triangular arbitrage over a prolonged period of time. Our results suggest that the foreign exchange market is internally self-consistent and provide a limited verification of market efficiency.
\end{abstract}


\section{Introduction}
\label{introduction}
The foreign exchange (FX) market is the world's largest financial market with an average daily trade volume of approximately $3.2$ trillion USD\footnote{The currency codes used throughout this paper are: USD - U.S. dollar, CHF - Swiss franc, JPY - Japanese yen, EUR - euro.}\cite{BIS} and liquidity\footnote{We consider the market to have high liquidity if there is a large depth of resting orders and this depth is refreshed quickly when orders are filled.} throughout the $24$ hour trading day. In this paper we focus on triangular arbitrage within the FX market. Triangular arbitrage represents one of the simplest arbitrage opportunities. However, there is, to our knowledge, no truly rigorous and robust study of triangular arbitrage in the finance literature. We believe the main reason for this to be the lack of availability of datasets with prices which are of sufficiently high-frequency \emph{and} which are also executable. As a result of the size and liquidity of the FX market price updates occur at extremely high frequencies\footnote{The EUR/USD rate has in excess of $100$ ticks a minute during the most liquid periods.} -- therefore one requires an equally high-frequency dataset to test for triangular arbitrage opportunities. In addition, it is necessary to know that the prices are ones at which a trade could indeed be executed, as opposed to simply being indicative price quotes. Our own datasets satisfy of both these criteria and hence enable us to carry out this study in a reliable way.

An \emph{indicative} bid/ask price\footnote{Bid/ask prices give the different prices at which one can buy/sell currency, with the ask price tending to be larger than the bid price. The exchange rate between EUR and USD may, for example, be quoted as $1.4085$/$1.4086$. A trader then looking to convert USD into EUR might have to pay $1.4086$ USD for each EUR, while a trader looking to convert EUR to USD may receive only $1.4085$ USD per EUR. The difference between the bid and ask prices is the bid-ask spread.} is a quote that gives an approximate price at which a trade can be executed; at a given time one may be able to trade at exactly this price or, as is often the case, the real price at which one executes the trade, the \emph{executable} price, differs from the indicative price by a few basis points\footnote{A basis point is equal to $1$/$100$th of a percentage point. In this paper we will also discuss points, where a point is the smallest price increment for an exchange rate. For example, for the EUR/JPY exchange rate, which takes prices of the order of $139.60$ over the studied period, $1$ point corresponds to $0.01$. In contrast, for the EUR/USD rate with typical values around $1.2065$, $1$ point corresponds to $0.0001$.}. The main purpose of an indicative price is to supply clients of banks with a gauge of where the price is. A large body of academic research into the FX market has been performed using indicative quotes often under the assumption that, due to reputational considerations, ``serious financial institutions'' are likely to trade at exactly the quoted price, especially if they are hit a short time after the quote is posted \cite{dacorogna,guillaume,dacorogna_book}. The efficiency of using indicative quote data for certain analyses has, however, been drawn into question \cite{martens,lyons}. In \cite{lyons}, Lyons highlights the fact that indicative prices are not transactable; that the indicative bid-ask spread, despite usually ``bracketing'' the actual tradeable spread, is usually two to three times as large; that during periods of high trading intensity market makers are too busy to update their indications; and that market makers themselves are unlikely to garner much of their high-frequency information from these indicative prices. In the FX market today indicative prices are typically updated by automated systems, nevertheless the quoted price is still not necessarily a price at which one could actually execute a trade.

Goodhart \textit{et al.} \cite{goodhart1995} performed a comparison of indicative bid-ask quotes from the Reuters FXFX page and executable prices from the Reuters D2000--2 electronic broking system, over a $7$ hour period, and found that the behaviour of the bid-ask spread and the frequency at which quotes arrived were quite different for the two types of quote. In particular, the spread from the D2000--2 system showed greater temporal variation, with the variation dependent upon the trading frequency. In contrast, the indicative price spread tended to cluster at round numbers, a likely artifact of their use as a market gauge. This discrepancy between indicative and executable prices is likely to be less important if one is performing a low frequency study, arguably down to time scales of $10$--$15$ minutes \cite{guillaume}. If, however, one is considering very high-frequency data, this difference becomes highly significant. For example, in \cite{goodhart1991} Goodhart and Figliuoli find a negative first-order auto-correlation in price changes at minute-by-minute frequencies using indicative data. In \cite{goodhart1995}, however, Goodhart finds no such negative auto-correlation when real transaction data is used. Indicative data seem particularly unsuitable to many market analyses today because banks are now able to provide their clients with automated executable prices through an electronic trading platform and so there is even less incentive for them to make their indicative quotes accurate.

Some analyses of triangular arbitrage have been undertaken using indicative data. In \cite{aiba}, Aiba \textit{et al.} investigate triangular arbitrage using quote data provided by information companies, for the set of exchange rates \{EUR/USD, USD/JPY, EUR/JPY\}, over a roughly eight week period in 1999. They find that, over the studied period, arbitrage opportunities appear to exist about 6.4\% of the time, or around 90 minutes each day, with individual arbitrages lasting for up to approximately $1,000$ seconds. In \cite{kollias}, Kollias and Metaxas investigate $24$ triangular arbitrage relationships, using quote data for seven major currencies over a one month period in 1998, and find that single arbitrages exist for some currency groups for over two hours, with a median duration of 14 and 12 seconds for the two transactions formed from \{USD/DEM, USD/JPY, DEM/JPY\}.

When considering whether triangular arbitrage transactions can be profitable, it is important to consider how long the opportunities persist. The time delay between identifying an opportunity and the arbitrage transaction being completed is instrumental in determining whether a transaction results in a profit because the price may move during this time interval. Kollias and Metaxas tested the profitability of triangular arbitrage transactions by considering execution delays of between $0$ and $120$ seconds and in a similar manner, Aiba \textit{et al.} accounted for delays by assuming that it took an arbitrageur between $0$ and $9$ seconds to recognize and execute an arbitrage transaction. Kollias and Metaxas found that, for some transactions, triangular arbitrage continued to be profitable for delays of $120$ seconds and Aiba \textit{et al.} for execution delays of up to $4$ seconds. These durations differ markedly from the beliefs of market participants and we suggest that this discrepancy results from the invalid use of indicative data in these studies.

In contrast to previous studies, in this paper, we use high-frequency, \emph{executable} price data to investigate triangular arbitrage. This means that, for each arbitrage opportunity identified, one could potentially have executed a trade at the arbitrage price. Furthermore, and importantly, we consider the issue of not completing an arbitrage transaction. In the FX market today, where electronic trading systems are widely used, it is possible to undertake the three constituent trades of an arbitrage transaction in a small number of milliseconds but, despite this execution speed, one is not guaranteed to complete an arbitrage transaction. We discuss the reasons for this.

The paper is organized as follows. In Section \ref{triarb} we define a triangular arbitrage and in Section \ref{data} describe the data used in this study. In Section \ref{properties} we investigate the properties of the triangular arbitrage opportunities and in Section \ref{profitability} consider the profitability of arbitrage transactions. In Section \ref{conclusions} we conclude.

\section{Triangular arbitrage}
\label{triarb}
Consider the situation where one initially holds $x_i$ euros. If one sells these euros and buys dollars, converts these dollars into Swiss francs and finally converts these francs into $x_f$ euros then if $x_f>x_i$ a profit is realized. This is a triangular arbitrage. Such opportunities should, in such a liquid market, be limited and if they do occur one would expect the difference $x_f-x_i$ to be extremely small. This then means that, when identifying arbitrage opportunities on a second-by-second time scale, the possible discrepancy between an indicative and an executable price becomes extremely important. It is, in fact, essential to use executable data if one is to draw reliable conclusions on whether triangular arbitrage opportunities exist.

Triangular arbitrage opportunities can be identified through the rate product
\begin{eqnarray}
\gamma(t)=\prod_{i=1}^{3}{r_i(t)},
\end{eqnarray}
where $r_i(t)$ denotes an exchange rate at time $t$ \cite{aiba}. An arbitrage is theoretically possible if $\gamma>1$, but a profit will only be realized if the transaction is completed at an arbitrage price.

For each group of exchange rates there are two unique rate products that can be calculated. For example, consider the set of rates \{EUR/USD, USD/CHF, EUR/CHF\}. If one initially holds euros, one possible arbitrage transaction is EUR$\rightarrow$USD$\rightarrow$CHF$\rightarrow$EUR with a rate product given by
\begin{eqnarray}
\gamma_1(t)=\bigg(\mathrm{EUR/USD_{bid}}(t)\bigg) . \bigg(\mathrm{USD/CHF_{bid}}(t)\bigg) . \bigg(\frac{1}{\mathrm{EUR/CHF_{ask}}(t)}\bigg).
\label{g1}
\end{eqnarray}
The second possible arbitrage transaction is EUR$\rightarrow$CHF$\rightarrow$USD$\rightarrow$EUR with a rate product
\begin{eqnarray}
\gamma_2(t)=\bigg(\frac{1}{\mathrm{EUR/USD_{ask}}(t)}\bigg) . \bigg(\frac{1}{\mathrm{USD/CHF_{ask}}(t)}\bigg) . \bigg(\mathrm{EUR/CHF_{bid}(t)}\bigg).
\label{g2}
\end{eqnarray}
These two rate products define all possible arbitrage transactions using this set of exchange rates.

\section{Data}
\label{data}
\begin{figure}[pb]
\centerline{\psfig{file=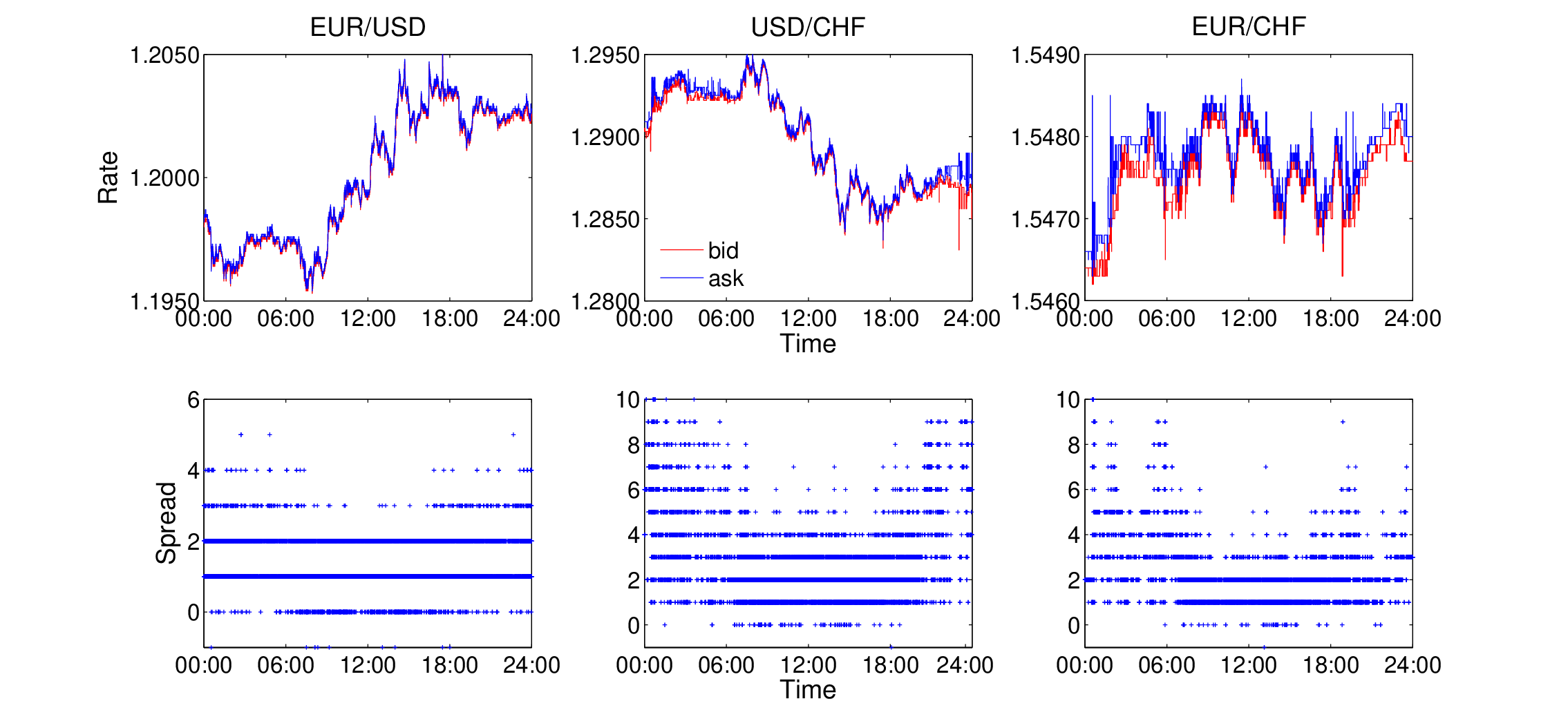,width=12.5cm}}
\vspace*{8pt}
\caption{Exchange rate time series for EUR/USD, USD/CHF and EUR/CHF on 10/12/2005. Upper: bid and ask prices. Lower: bid-ask spread. Each marker represents the spread at a single time-step. The vertical axes have been truncated to make the detail around the typical values clearer.}
\label{CEU_rates_eg}
\end{figure}

The data used for the analysis was provided by HSBC bank, one of the largest FX market-making banks in the world, and consists of second-by-second executable prices for \{EUR/USD, USD/CHF, EUR/CHF, EUR/JPY, USD/JPY\}. Triangular arbitrage opportunities are investigated for the transactions involving \{EUR/USD, USD/CHF, EUR/CHF\} and \{EUR/USD, USD/JPY, EUR/JPY\} for all week days over the period 10/02/2005--10/27/2005 and the results obtained compared with those for the two earlier periods: 10/27/2003--10/31/2003 and 10/01/2004--10/05/2004.\footnote{All times in this paper are given in GMT. The full day 10/28/2005 is excluded from the analysis for the JPY group of exchange rates due to an error with the data feed on this day. During periods of lower liquidity it is possible that there were times at which no party was offering a bid and/or ask price. At these times it would not have been possible to complete a triangular transaction involving the missing exchange rate and so the associated rate product is simply set to zero.} The full data set consists of approximately $2.6$ million data points for each of the rate products $\gamma_1$ and $\gamma_2$, $5.2$ million data points for each of the currency groups and $10.4$ million data points in total. A rate product, indicating whether or not a triangular arbitrage opportunity existed, was found for each of these points. A sample of one of the sets of exchange rates and the corresponding time series of bid-ask spreads is shown in Fig. \ref{CEU_rates_eg}.

\section{Arbitrage properties}
\label{properties}
\subsection{Rate products}
\label{rateproducts}
\label{analysis:rp}
\begin{figure}[pb]
\centerline{\psfig{file=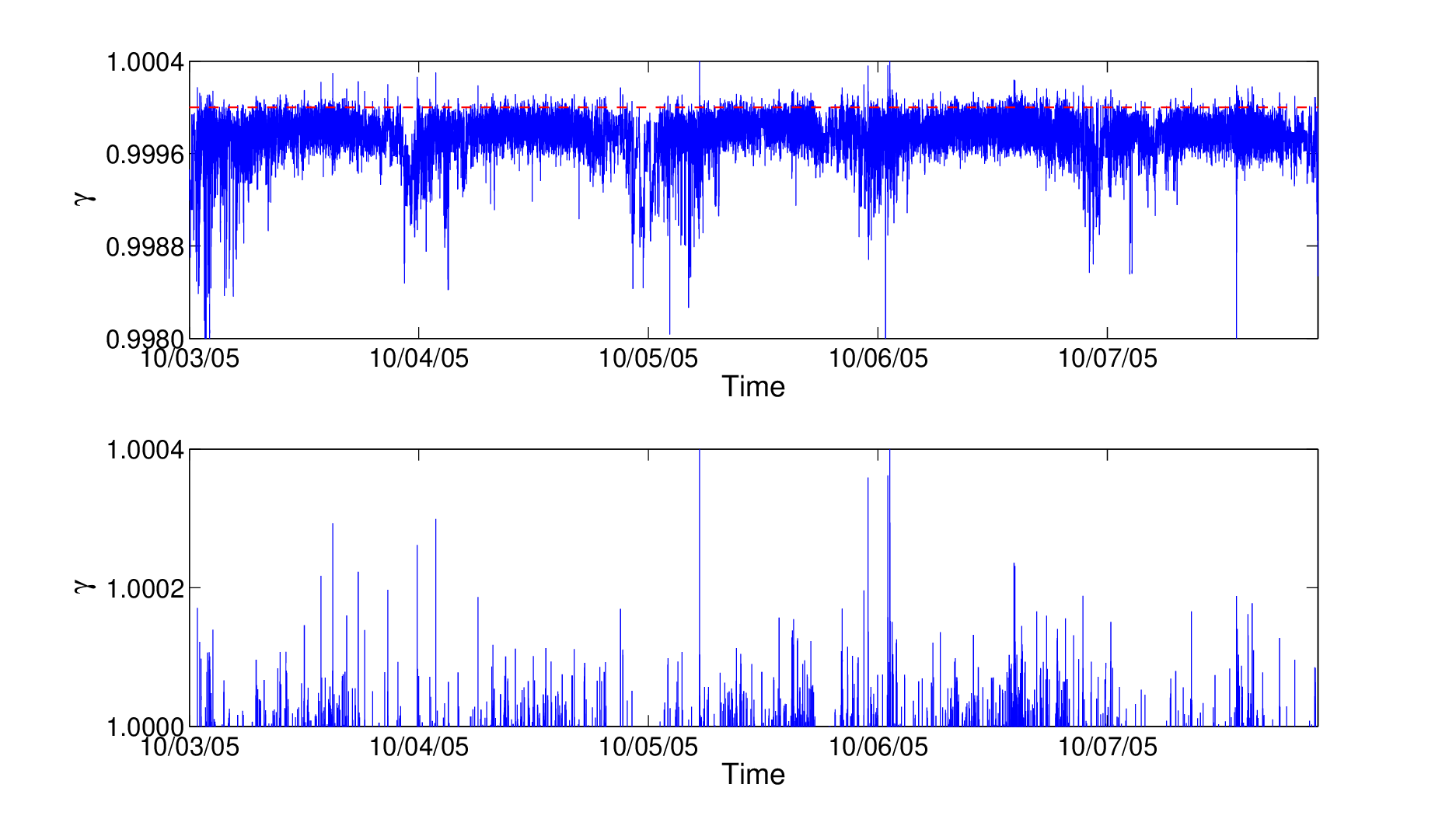,width=12.5cm}}
\vspace*{8pt}
\caption{Rate product evolution for the period 10/03/2005--10/07/2005 for the transaction EUR${\rightarrow}$USD${\rightarrow}$JPY${\rightarrow}$EUR. Upper: all rate products, with a few extreme values removed so that the structure around the typical values is clearer. All points above the red line correspond to potential triangular arbitrages. Lower: the same plot truncated vertically at $\gamma=1$ so that each spike represents an arbitrage opportunity.}
\label{rp_eg}
\end{figure}

Figure \ref{rp_eg} shows an example of the temporal evolution of the rate product, $\gamma$, over one of the weeks analyzed. If it were possible to buy and sell a currency at exactly the same price then one would expect the rate product to always equal one. However, the prices at which currencies can be bought and sold differ, with the ask price exceeding the bid price and, as a result, the rate product is typically expected to be slightly less than one. Rate products with a value just below one can be considered to fall in a region of triangular parity\footnote{Triangular parity implies that the direct exchange rate is equal to the exchange rate generated through the cross-rates. For example, EUR/USD = (EUR/JPY)/(USD/JPY), where one needs to use the correct bid and ask price to construct the synthetic exchange rate.}.

\begin{figure}[pb]
\centerline{\psfig{file=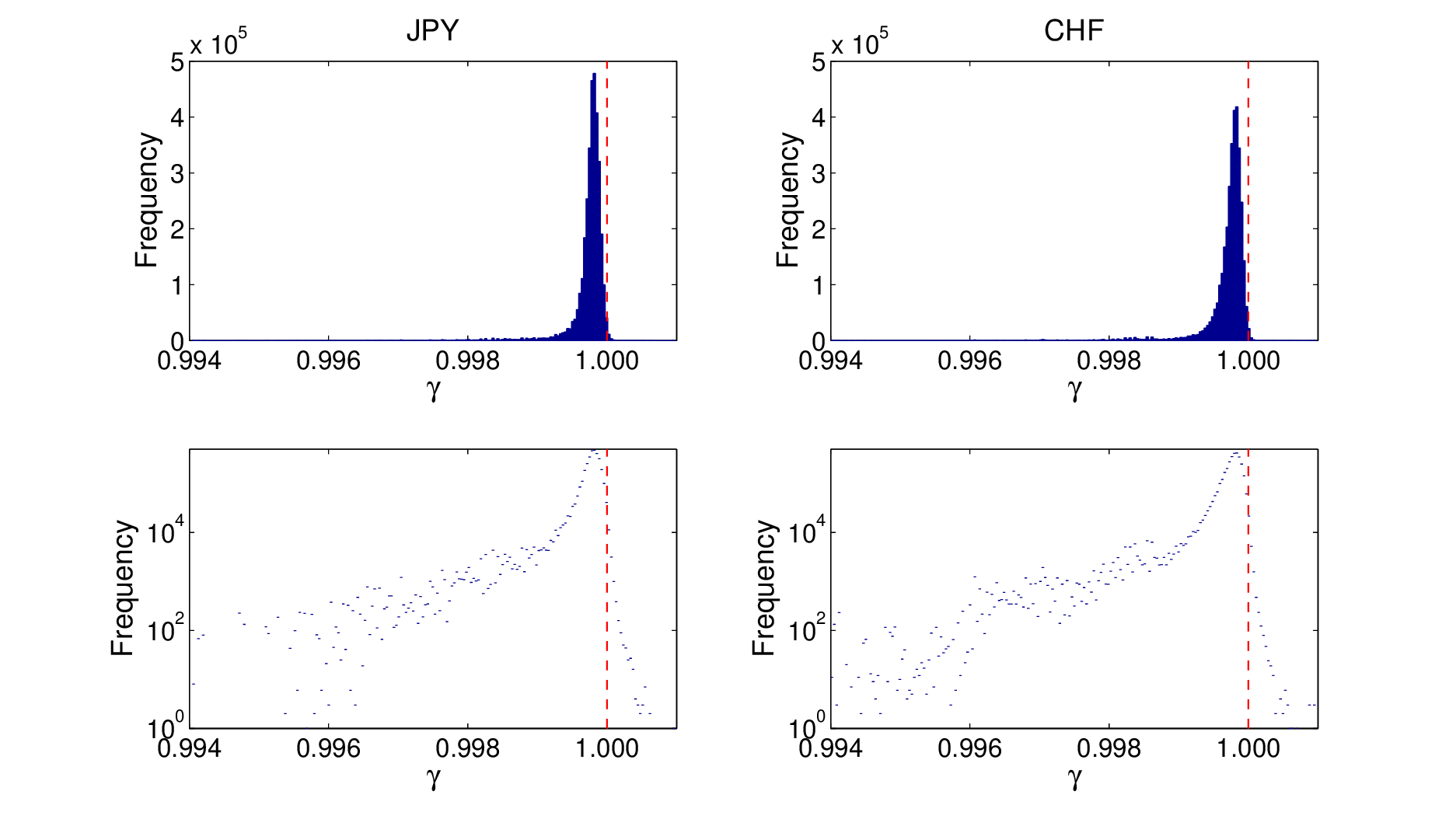,width=12.5cm}}
\vspace*{8pt}
\caption{Occurrence frequency for rate products of different magnitudes for the period 10/02/2005--10/27/2005. Upper: aggregated results for both JPY transactions and CHF transactions. Any parts of the histograms to the right of the line at $\gamma=1$ correspond to potential triangular arbitrages. The JPY panels show all data points within this period and the CHF panels all points except a few at very small and very large $\gamma$. Lower: the same distributions on a log-linear scale.}
\label{rp_full}
\end{figure}

The distributions in Fig. \ref{rp_full} show that, as expected, the rate product does tend to be slightly less than one and typically $\gamma\in[0.9999,1)$. The log-linear plots also highlight that the distributions possess long tails extending to smaller values of the rate product and that there are some times when $\gamma>1$. This means that, for the majority of deviations from triangular parity, the individual exchange rates are shifted in such a direction that triangular arbitrage is not possible, but that occasionally potential profit opportunities do occur. Over the four week period analyzed, there are $10,018$ triangular arbitrage opportunities for the two CHF-based transactions given by Eqs. (\ref{g1}) and (\ref{g2}) and $11,367$ for the equivalent JPY transactions. We now establish both the duration and magnitude of these potential arbitrages and attempt to determine whether or not they represent genuine, executable profit opportunities.

\subsection{Durations}
\label{durations}
Firstly, we consider the length of periods for which $\gamma>1$ and thus over which triangular arbitrage opportunities exist. For an arbitrage of $X$ seconds, $\gamma>1$ for more than $X-1$, but less than $X$ consecutive seconds. The summary statistics in Table \ref{table:arb_length} demonstrate that the vast majority of arbitrage opportunities are very short in duration. Although some opportunities appear to exist for in excess of $100$s, for both currency groups 95\% last for $5$ seconds or less and 60\% for $1$ second or less.

\begin{table}[ht]
\tbl{\label{table:arb_length}Summary statistics for the duration of arbitrage opportunities for the two JPY and two CHF transactions for the period 10/02/2005--10/27/2005. An opportunity labelled as $X$s lasted for more than $X-1$ but less than $X$ seconds.}
{\begin{tabular}{@{}cccccccccccc@{}}
\arrayrulecolor{black}
\toprule
\multirow {2}*{Transaction}\hphantom{000} & \multicolumn{4}{c}{Duration (s)}\hphantom{00}
& &\multicolumn{6}{c}{Percentage of opportunities}\\
\cline{2-5}\cline{7-12}
&mean & median & min. & max. &\hphantom{00} & 1s & 2s & 3s & 4s &5s & $>5$s \\
\hline
JPY\hphantom{000} &2.01 &1 &1 &70 &\hphantom{00} &60 &21 &8 &4 &2 &5\\
CHF\hphantom{000} &2.09 &1 &1 &144 &\hphantom{00} &60 &21 &8 &4 &2 &5\\
\botrule
\end{tabular}}
\end{table}
The three constituent trades of a triangular arbitrage transaction can be submitted extremely fast using an electronic trading system, but there is still a delay from the time that the opportunity is identified, and the trades initiated, to the time that the trades arrive at the price source. Although this delay is typically only of the order of milliseconds, it is nonetheless significant. If the trader places each trade as a limit order, that will only be filled at the arbitrage price, then if one of the prices moves, due to trading activity or the removal of a price by the party posting it, then the transaction will not be completed. For example, consider the transaction EUR$\rightarrow$USD$\rightarrow$CHF$\rightarrow$EUR and assume that a trader completes the EUR$\rightarrow$USD and CHF$\rightarrow$EUR transactions at arbitrage prices. If the USD$\rightarrow$CHF transaction is not completed, because the USD/CHF has moved to an arbitrage-free price, the trader will be left with a long position in USD and a short position in CHF. The trader may choose to unwind this position immediately by converting USD into CHF and this transaction will cost the amount by which the price has moved from the arbitrage price. Over a short time-scale, this is likely to be $1$--$2$ points (approximately $1.5$--$2$ basis points). Incomplete arbitrage transactions therefore typically cost a small number of basis points.

The extremely short time scales involved in these trades  means that the physical distance between the traders and the location where their trades are filled is important in determining which trade arrives first and is completed at the arbitrage price. This explains why a number of exchanges have begun to offer the possibility of locating trading systems on their premises. This is known as co-location.

A trader has a higher chance of completing an arbitrage transaction for opportunities with longer durations because the arbitrage prices remain active in the market for longer. When an arbitrage signal is received, however, there is no way of knowing in advance how long the arbitrage will exist for. Over half of all arbitrage opportunities last for less than $1$ second and so there is a high probability that any signal that is traded on is generated by an opportunity of less than a second. This includes many opportunities that last for only a few milliseconds. For these opportunities there is a smaller chance of the transaction being completed at an arbitrage price. For each attempted arbitrage, one cannot eliminate the risk that one of the prices will move to an arbitrage-free price before the transaction is completed.

\subsection{Magnitudes}
\label{sub_arb_mag}
Given these risks, one possible criterion that could be used, in order to decide whether or not to trade, is the magnitude of the apparent opportunity. If the value of the rate product is large, and thus it appears that a significant profit could potentially be gained, one may decide that the potential reward outweighs the associated risks and execute the arbitrage transactions. In this section we consider the magnitudes of the arbitrage opportunities.
\begin{table}[ht]
\tbl{\label{bp_analysis}The number and mean duration of arbitrage opportunities exceeding different thresholds for the two JPY transactions and two CHF transactions for the period 10/02/2005--10/27/2005. A one basis point threshold corresponds to a rate product of $\gamma\geq1.0001$.}
{\begin{tabular}{@{}p{0.6cm}p{2.95cm}llllllllllll@{}}
\arrayrulecolor{black}
\toprule
&Basis point threshold &0 &0.5 &1 &2 &3 &4 &5 &6 &7 &8 &9 &10\\
\hline
\multirow{2}{*}{JPY} &No. of arbitrages &17,314 &5,657 &1,930 &220 &50 &21 &7 &3 &1 &1 &1 &0\\
&Mean duration (s) &3.3 &3.0 &2.6 &1.5 &1.6 &1.4 &1.6 &1.0 &1.0 &1.0 &1.0 &0\\
\hline
\multirow{2}{*}{CHF} &No. of arbitrages &10,018 &2,376 &649 &119 &37 &20 &15 &7 &6 &6 &6 &5\\
&Mean duration (s) &2.1 &1.5 &1.5 &1.9 &1.9 &1.8 &2.0 &2.6 &2.8 &2.8 &2.3 &2.2\\
\botrule
\end{tabular}}
\end{table}

Table \ref{bp_analysis} demonstrates that most arbitrage opportunities have small magnitudes, with $94$\% less than $1$ basis point for both the JPY and CHF. An arbitrage opportunity of $1$ basis point corresponds to a potential profit of $100$ USD on a $1$ million USD trade. A single very large trade, or a large number of smaller trades, would thus be required in order to realize a significant profit on such an opportunity. Large volume trades are, however, often not possible at the arbitrage price. Consider the transaction EUR${\rightarrow}$USD${\rightarrow}$JPY${\rightarrow}$EUR at a time when EUR/USD$_{\mathrm{bid}}=1.2065$, USD/JPY$_{\mathrm{bid}}=115.72$ and EUR/JPY$_{\mathrm{ask}}=139.60$, resulting in $\gamma=1.000115903$. If there are only $10$ million EUR available on the first leg of the trade at an arbitrage price, then the potential profit is limited to $1,159$ EUR. In practice, the amount available at the arbitrage price may be substantially less than $10$ million USD and consequently the potential profit correspondingly smaller.

This calculation also assumes that it is possible to convert the full volume of currency at an arbitrage price for each of the other legs of the transaction. In practice, however, the volumes available on these legs will also be limited. For example, again consider the case where there are $10$ million EUR available at an arbitrage price on the first leg of the above transaction. If the full $10$ million are converted into USD, the trader will hold  $12.065$ million USD. There may, however, only be $10$ million USD available at an arbitrage price on the next leg of the trade. In order for the full volume to be traded at an arbitrage price, the trader should therefore limit the initial EUR trade to $10/1.2065=8.29$ million EUR. The volume available on the final leg of the trade would also need to be considered in order to determine the total volume that can be traded at an arbitrage price. This volume and the total potential profit are therefore determined by the leg with the smallest available volume.

Occasionally, larger magnitude arbitrage opportunities can occur. Table \ref{bp_analysis} shows that, over the studied period, there are potential arbitrages of more than $9$ basis points for both currency groups, with a mean duration\footnote{Each mean duration represents an upper bound. This is because each opportunity labelled as $X$s may have existed for anywhere between $X-1$ and $X$ seconds, but in calculating the mean duration we assume that it lasted for exactly $X$ seconds.} of in excess of $2$ seconds for the large CHF opportunities. This duration suggests that one would have stood a good chance of completing an arbitrage transaction for one of these opportunities. This mean, however, was calculated using only six opportunities and so does not represent a reliable estimate of the expected duration. The fact that these large opportunities occur so infrequently (with only around $20$ potential arbitrages in excess of $4$ basis points occurring for each transaction over the four week period analyzed) means that trading strategies that only trade on these larger opportunities would need to make large volume trades in order to realize significant profits. As discussed above, though, the volume available at the arbitrage price is always limited.

\subsection{Seasonal variations}
\label{seasonalities}
\begin{figure}[pb]
\centerline{\psfig{file=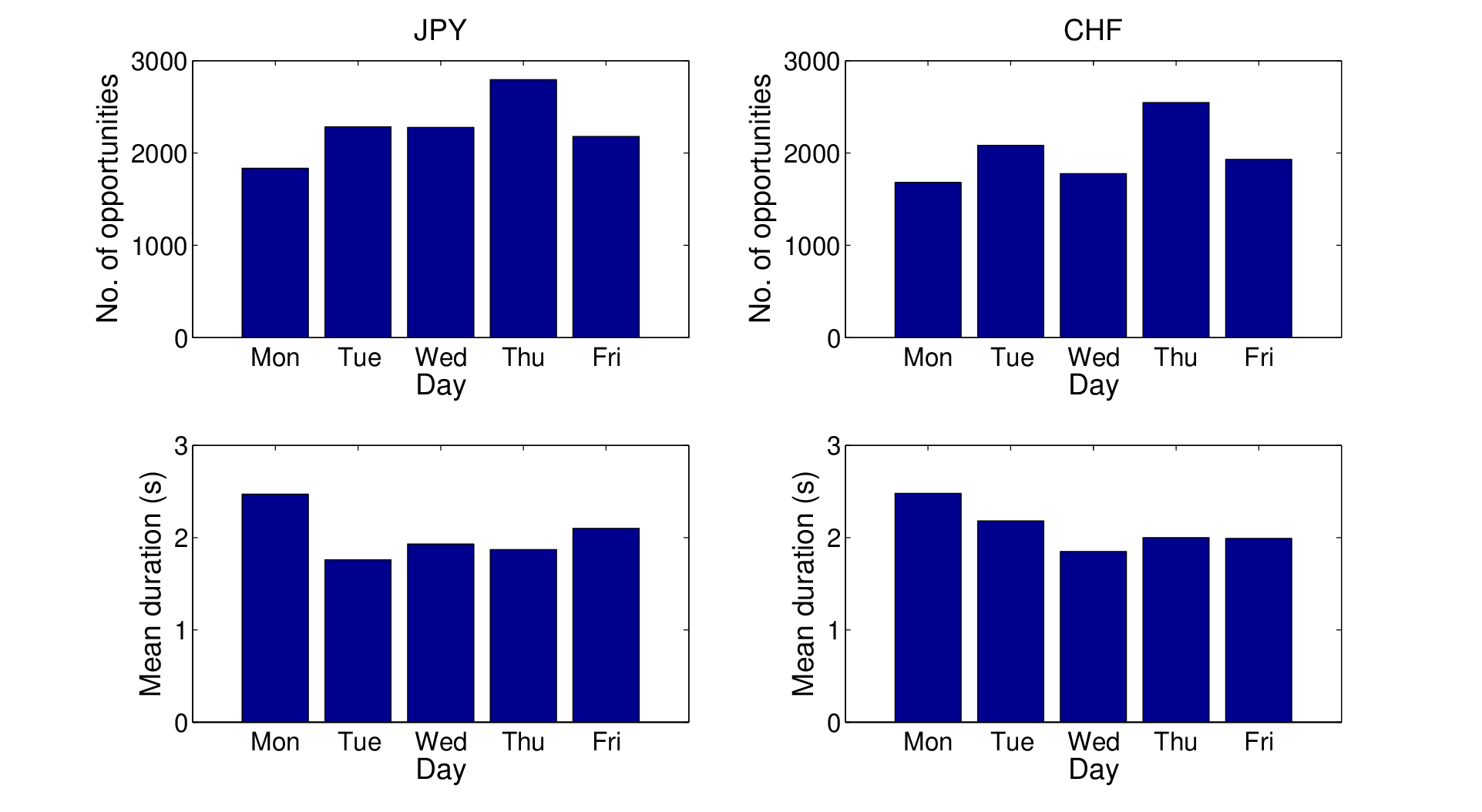,width=12.5cm}}
\vspace*{8pt}
\caption{Daily arbitrage statistics for the period 10/02/2005--10/27/2005. Upper: the number of arbitrage opportunities. Lower: mean duration of arbitrage opportunities.}
\label{daily_distributions}
\end{figure}

\begin{figure}[pb]
\centerline{\psfig{file=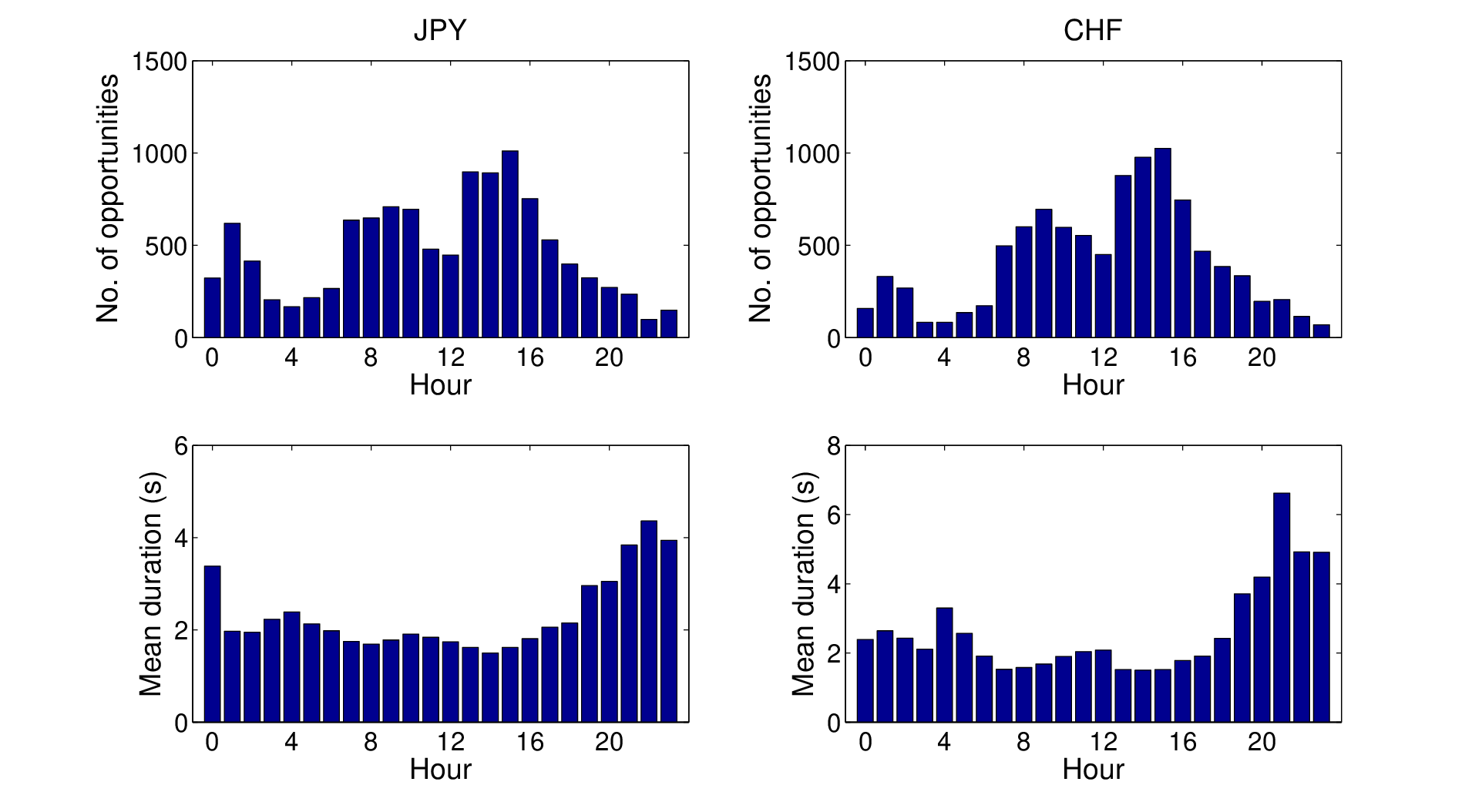,width=12.5cm}}
\vspace*{8pt}
\caption{Hourly arbitrage statistics for the period 10/02/2005--10/27/2005. Upper: the number of arbitrage opportunities. Lower: mean duration of arbitrage opportunities.}
\label{hourly_distributions}
\end{figure}

We now consider whether there is any seasonality in the number and duration of arbitrage opportunities by investigating daily and hourly statistics. Figure \ref{daily_distributions} shows that the number of arbitrage opportunities per day, and their mean duration, is reasonably uniform across days. Fig. \ref{hourly_distributions}, however, demonstrates that there is a large amount of variation in these quantities for different hours of the day. Both the JPY and CHF transactions show a particularly small number of opportunities, with a large mean duration, between approximately 22:00 and 01:00, and a large number of opportunities, with a short duration, between 13:00 and 16:00. In general, the hours with larger number of arbitrage opportunities correspond to those with shorter mean durations and \emph{vice-versa}.

\begin{table}[ht]
\tbl{\label{market_hours}Grey blocks indicate the hours corresponding to high liquidity for the Asian, European and American markets.}
{\begin{tabular}{@{}|p{1.2cm}|p{0.12cm}|p{0.12cm}|p{0.12cm}|p{0.12cm}|p{0.12cm}|p{0.12cm}|p{0.12cm}|p{0.12cm}|p{0.12cm}|p{0.12cm}|p{0.12cm}|p{0.12cm}|p{0.12cm}|p{0.12cm}|p{0.12cm}|p{0.12cm}|p{0.12cm}|p{0.12cm}|p{0.12cm}|p{0.12cm}|p{0.12cm}|p{0.12cm}|p{0.12cm}|p{0.12cm}|@{}}
\arrayrulecolor{black}
\hline
&0 &1 &2 &3 &4 &5 &6 &7 &8 &9 &10 &11 &12 &13 &14 &15 &16 &17 &18 &19 &20 &21 &22 &23\\
\hline
Asia &\cellcolor[gray]{0.8} &\cellcolor[gray]{0.8} &\cellcolor[gray]{0.8} &\cellcolor[gray]{0.8} &\cellcolor[gray]{0.8} &\cellcolor[gray]{0.8} &\cellcolor[gray]{0.8} &\cellcolor[gray]{0.8} &\cellcolor[gray]{0.8} &\cellcolor[gray]{0.8} &\cellcolor[gray]{0.8} & & & & & & & & & & & & & \\
\hline
Europe & & & & & & & &\cellcolor[gray]{0.8} &\cellcolor[gray]{0.8} &\cellcolor[gray]{0.8} &\cellcolor[gray]{0.8} &\cellcolor[gray]{0.8} &\cellcolor[gray]{0.8} &\cellcolor[gray]{0.8} &\cellcolor[gray]{0.8} &\cellcolor[gray]{0.8} &\cellcolor[gray]{0.8} &\cellcolor[gray]{0.8} & & & & & & \\
\hline
Americas & & & & & & & & & & & & & &\cellcolor[gray]{0.8} &\cellcolor[gray]{0.8} &\cellcolor[gray]{0.8} &\cellcolor[gray]{0.8} &\cellcolor[gray]{0.8} &\cellcolor[gray]{0.8} &\cellcolor[gray]{0.8} &\cellcolor[gray]{0.8} &\cellcolor[gray]{0.8} &\cellcolor[gray]{0.8} &\cellcolor[gray]{0.8}\\
\hline
\end{tabular}}
\end{table}

These differences can be explained by the variation in liquidity throughout the trading day. Table \ref{market_hours} shows the periods during which the Asian, European and American FX markets are at their most liquid. The period of highest liquidity is from 08:00--16:00; over almost all of this period two of the markets are highly liquid at similar time. The period of least liquidity is from around 22:00--01:00. The hours with the largest number of arbitrage opportunities and the shortest mean durations in Fig. \ref{hourly_distributions}, thus correspond to the periods of highest liquidity. This observation of more arbitrage opportunities during the periods of highest liquidity seems counter-intuitive, but can be explained as follows. During liquid periods the bid-ask spread is narrower (see Fig. \ref{CEU_rates_eg}) and prices move around at a higher frequency due to the large volume of trading. This results in more price mis-alignments and thus more potential arbitrages. The high trade frequency, however, also ensures that the mis-pricings are quickly traded away or removed and thus that any arbitrage opportunities are short-lived. In contrast, during less liquid periods the spread is wider, and the trading volume lower, leading to fewer arbitrage opportunities. The smaller number of traders available to correct any mis-pricings during less liquid times also results in the arbitrages having a longer duration.

\subsection{Annual variations}
The analysis so far has focused on a four week period in October 2005. In this section we consider how the number and distribution of triangular arbitrage opportunities has changed over the years by comparing results for three typical weeks between 2003 and 2005: 10/27/2003--10/31/2003, 11/01/2004--11/05/2004 and 10/17/2005--10/21/2005. These three weeks all fall at the same time of year and so any seasonal factors are eliminated.

\begin{table}[ht]
\tbl{\label{table:length_comp}Comparison of the number and percentage of arbitrage opportunities of selected durations and the mean and standard deviation of the rate product probability distributions for the periods 10/27/2003--10/31/2003, 11/01/2004--11/05/2004 and 10/17/2005--10/21/2005. An opportunity labelled as $X$s lasted for more than $X-1$ but less than $X$ seconds.}
{\begin{tabular}{@{}cccccccccccc@{}}
\arrayrulecolor{black}
\toprule
\multirow {2}*{Transaction} &\multirow {2}*{Year} &\multirow {2}*{No. arbitrages} &\multicolumn{6}{c}{Percentage of opportunities} & &\multicolumn{2}{c}{Rate product statistics}\\
\cline{4-9}\cline{11-12}
& & &1s &2s &3s &4s &5s &$>5s$ & &mean &stand. dev.\\
\hline
&2003 &$4,220$ &40 &30 &14 &6 &3 &7 & &0.999625 &$4.32\times10^{-4}$\\
JPY &2004 &$3,662$ &49 &28 &12 &5 &3 &3 & &0.999723 &$2.25\times10^{-4}$\\
&2005 &$2,963$ &62 &21 &7 &4 &3 &3 & &0.999758 &$2.17\times10^{-4}$\\
\hline
&2003 &$3,590$ &41 &29 &13 &6 &4 &7 & &0.999549 &$6.02\times10^{-4}$\\
CHF &2004 &$3,441$ &49 &27 &11 &5 &3 &5 & &0.999663 &$3.54\times10^{-4}$\\
&2005 &$2,672$ &64 &20 &8 &3 &1 &4 & &0.999725 &$3.10\times10^{-4}$\\
\botrule
\end{tabular}}
\end{table}
\begin{figure}[pb]
\centerline{\psfig{file=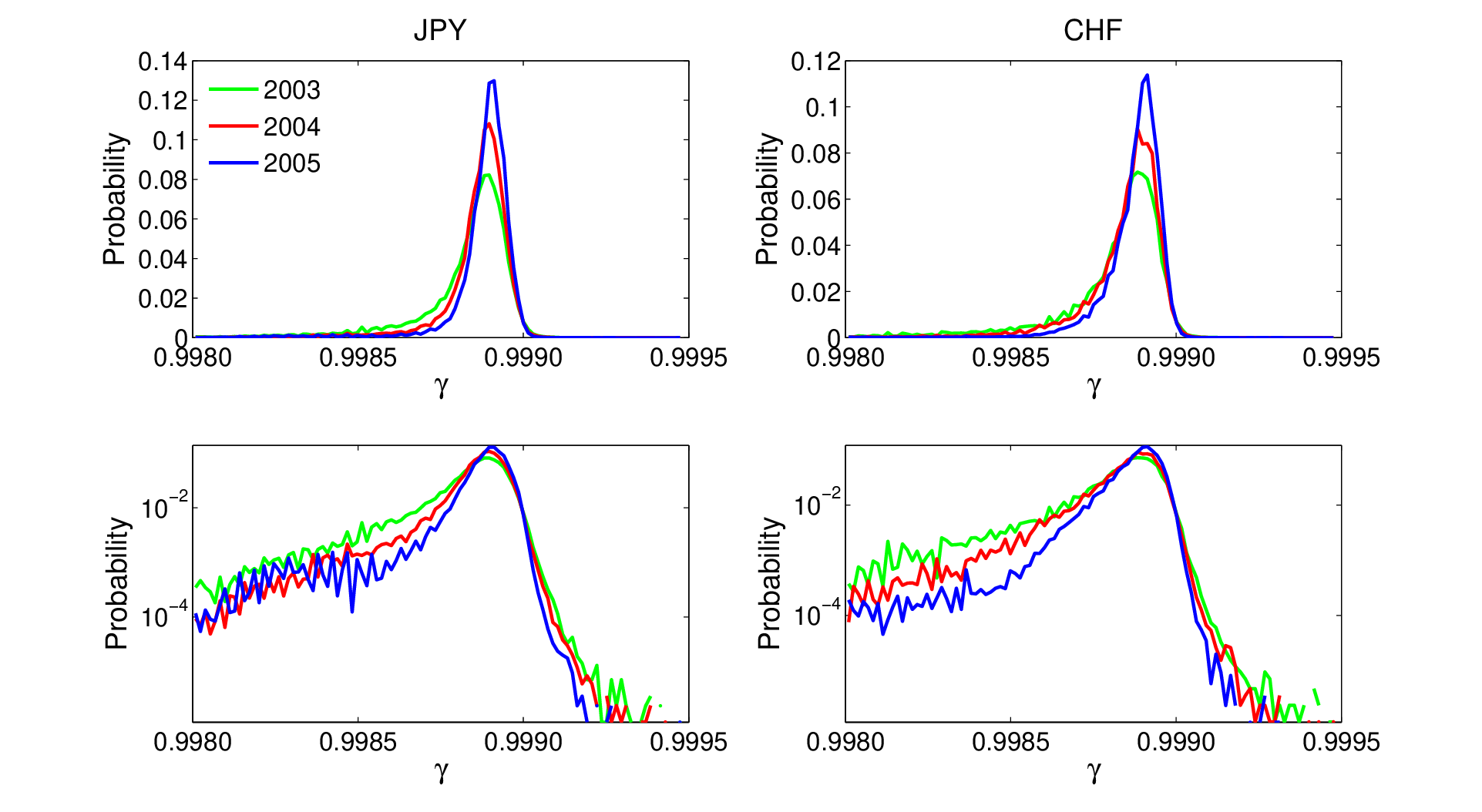,width=12.5cm}}
\vspace*{8pt}
\caption{Comparison of the rate product probability distributions for the periods 10/27/2003--10/31/2003, 11/01/2004--11/05/2004 and 10/17/2005--10/21/2005. Lower: distributions on a log-linear scale.}
\label{time_comp_rate_products}
\end{figure}

Table \ref{table:length_comp} shows that the number of arbitrage opportunities decreased from 2003--2005 for the JPY and CHF transactions. This can be explained by the increasingly wider use of electronic trading platforms and trading algorithms over this period. These systems enabled traders to execute trades faster and to react more quickly to price changes, which in turn gave rise to increased trading efficiency, fewer mis-pricings and fewer triangular arbitrage opportunities. Table \ref{table:length_comp} also demonstrates the significant effect that this increased execution speed had on the duration or arbitrage opportunities. From 2003--2005, the proportion of opportunities lasting less than $1$ second increased from 40\% to 62\% for the JPY transactions and from 41\% to 64\% for the CHF transactions and the proportion of opportunities lasting in excess of $5$ seconds halved for both sets of transactions.

The probability distributions in Fig. \ref{time_comp_rate_products} and the distribution statistics in Table \ref{table:length_comp} provide further evidence of the increased pricing efficiency of the FX market from 2003 to 2005. Over this period the distribution of rate products becomes concentrated in a sharper peak, with a smaller standard deviation and mean closer to one, demonstrating that triangular parity holds a larger proportion of the time.

\section{Profitability}
\label{profitability}
Finally, we provide further insights into the profitability of trading on triangular arbitrage signals by running simulations to determine the profit or loss that could potentially be achieved using different trading strategies. For the full time series of JPY and CHF rate products, over the period 10/02/2005--10/27/2005, we execute a simulated trade each time $\gamma$ exceeds some threshold amount $\gamma_t$. We consider the cases $\gamma_t=1$, i.e. all arbitrage signals are traded on irrespective of their magnitude, and $\gamma_t=1.00005$ and $1.0001$, corresponding to thresholds of half and one basis points respectively. The following two scenarios are considered for determining whether an arbitrage is filled:
\begin{itemize}
\item[(1)]Each traded arbitrage is filled with a fixed probability $p_1$.
\item[(2)]All arbitrages with a duration $\ell\geq1$ second are definitely filled. All opportunities traded on with a length $\ell<1$ second are filled with probability $p_2$.
\end{itemize}
For each completed arbitrage transaction, a profit determined by the rate product at the corresponding time step is received and for each unfilled transaction a fixed loss, $\lambda$, is incurred.\footnote{A fixed loss for each unfilled transaction is unrealistic and means that it is not possible to reliably estimate the volatility of the returns. It is, however, a reasonable first approximation.} We assume that each arbitrage opportunity with a duration $\ell\geq1$ second can only be traded on once, at its initial value, because if the simulated trader is left unfilled a competing trader must have been filled, resulting in the opportunity being removed. It is further assumed that, for each filled transaction, there is sufficient liquidity on each leg of the trade for it to be fully completed at the arbitrage price.

\begin{figure}[pb]
\centerline{\psfig{file=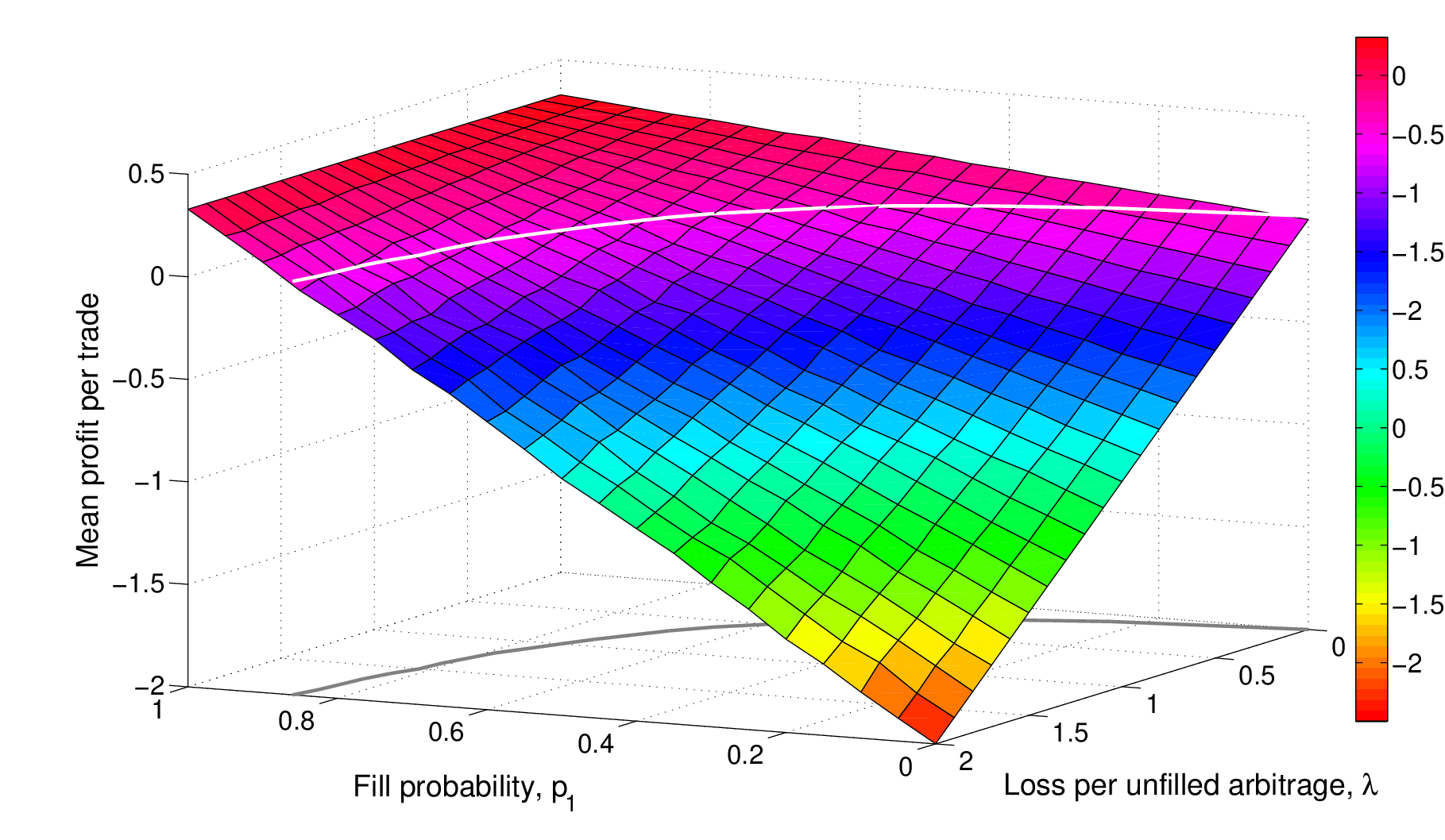,width=12.5cm}}
\vspace*{8pt}
\caption{Mean profit/loss per trade (in basis points) as a function of the probability of a transaction being filled at an arbitrage price and the loss incurred on missed arbitrages for JPY transactions over the period 10/02/2005--10/27/2005. A trade threshold $\gamma=1$ and scenario (1) are assumed. The white curve and its projection show the break-even fill probabilities. The probabilities are averaged over $100$ simulations.}
\label{plsurface}
\end{figure}

Figure \ref{plsurface} shows the mean profit per trade for scenario (1), as a function of $p_1$ and $\lambda$, for the JPY transactions. For a typical fixed loss per unfilled arbitrage of $\lambda=1.5$ (see Section \ref{durations}), an $80$\% fill probability is required to just break-even. Even for $p_1=1$, the maximum potential profit is less than half a basis point per transaction (about $50$ USD on a $1$ million USD trade).

\begin{figure}[pb]
\centerline{\psfig{file=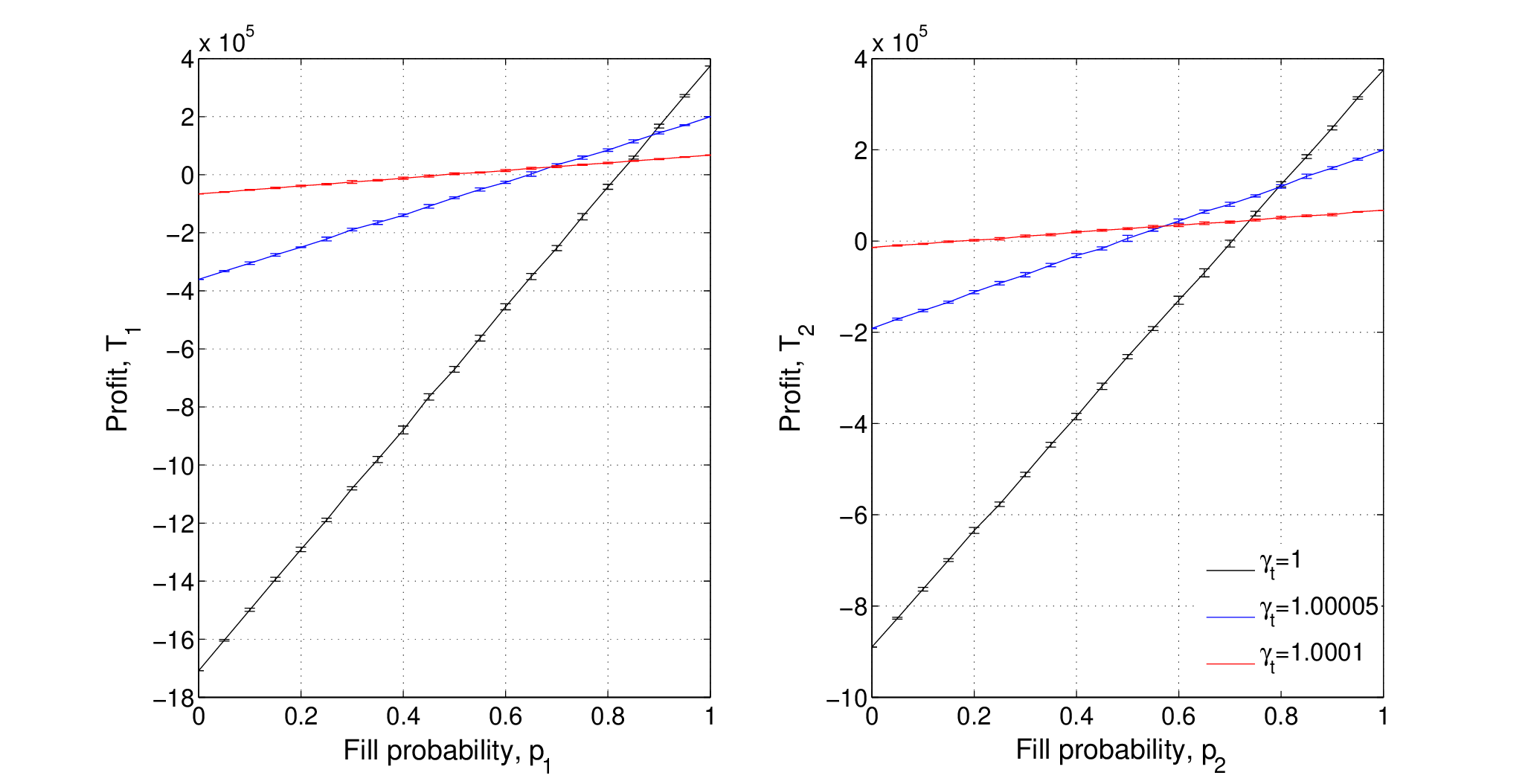,width=12.5cm}}
\vspace*{8pt}
\caption{Total profit (in EUR) for JPY transactions over the period 10/02/2005--10/27/2005. Each arbitrage transaction is traded with an initial currency outlay of $1$ million EUR and each completed transaction is filled for the full traded volume. We assume a fixed loss $\lambda=1.5$ basis points for each incomplete arbitrage transaction. Left: scenario (1). Right: scenario (2). Error bars indicate the standard deviation in the profit over $100$ simulations.}
\label{profits}
\end{figure}

We consider the total potential profit for the JPY transactions over the four week period 10/02/2005--10/27/2005 by simulating a trade of $1$ million EUR, each time $\gamma>\gamma_t$, and assuming a loss of $\lambda=1.5$ basis points for each incomplete arbitrage transaction. Figure \ref{profits} shows that for a 100\% fill probability, and a trade threshold of $\gamma_t=1$, a total profit of just under $400,000$ EUR appears possible for both scenarios (1) and (2). For higher values of $\gamma_t$, and a 100\% fill probability, the potential profit over the same period is smaller. The profit is smaller for higher $\gamma_t$ because there are fewer opportunities exceeding the thresholds and so fewer profit opportunities. The larger mean profit possible for each opportunity exceeding $\gamma_t$ is not sufficient to compensate for their reduced frequency. For a fill probability of zero, the lower trade frequency at higher thresholds limits the total possible loss relative to lower thresholds.

In order to achieve the $400,000$ EUR profit, it would have been necessary to stake $1$ million EUR more than $17,000$ times. If we estimate transaction fees and settlement costs at $2$ EUR per trade, then each arbitrage transaction costs $6$ EUR. The total cost of $17,000$ transactions is then $102,000$ EUR, which is a significant proportion of the potential profits. This profit is also likely to be a significant over-estimate. In the simulations, we assumed that each arbitrage transactions is completed for the full $1$ million EUR initially staked. As discussed in Section \ref{sub_arb_mag}, however, the amount available at the arbitrage price is limited and may be less than this amount. More importantly, a 100\% fill probability is extremely unrealistic and in practice the achievable fill probability will be significantly smaller. At a still unrealistic fill probability of $p_2=0.8$, for scenario (2), the potential profit is reduced to around $100,000$ EUR. This potential profit is already very similar to the estimated transaction costs and there are additional infrastructure costs that also need to be considered.

\begin{figure}[pb]
\subfigure{\psfig{file=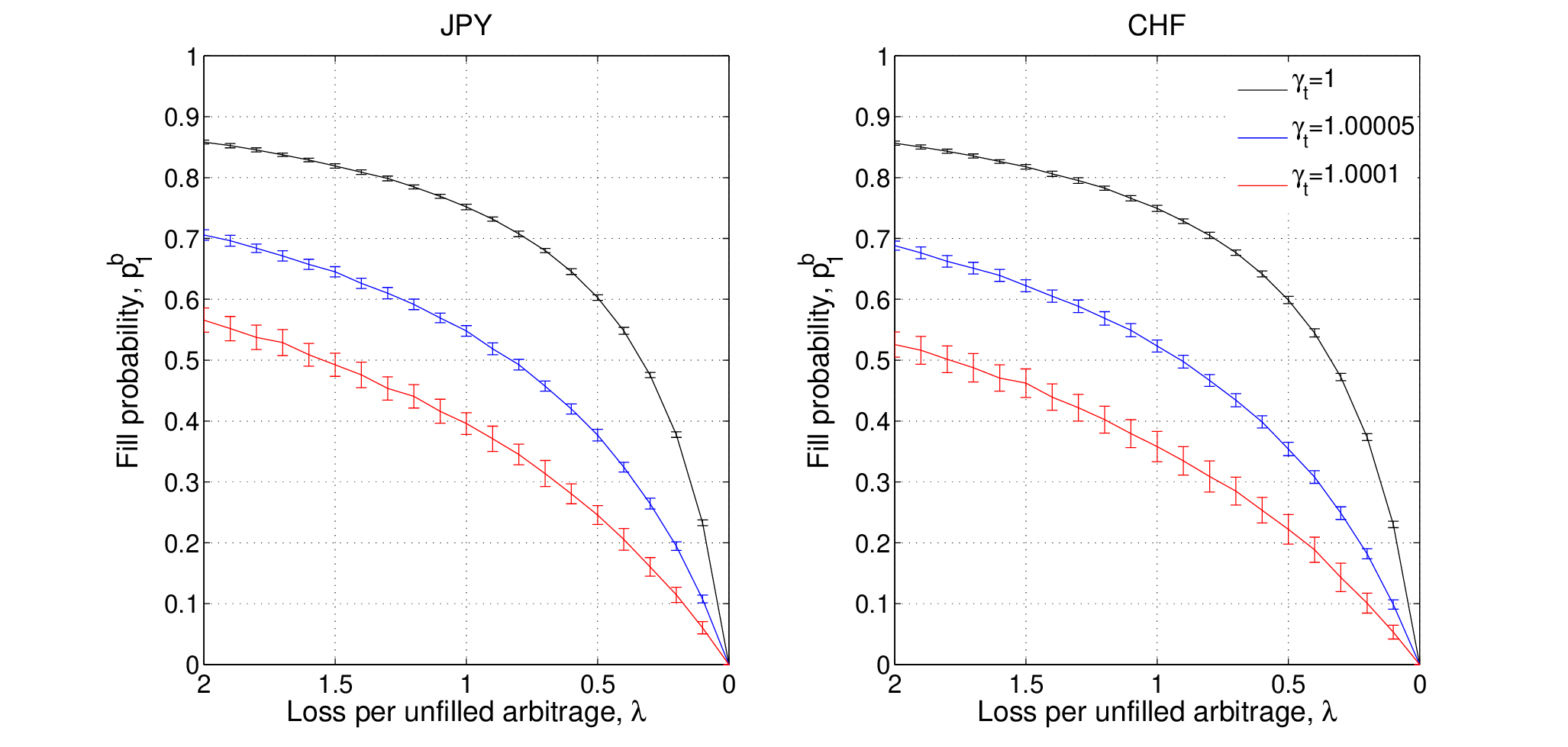,width=12.5cm}}
\subfigure{\psfig{file=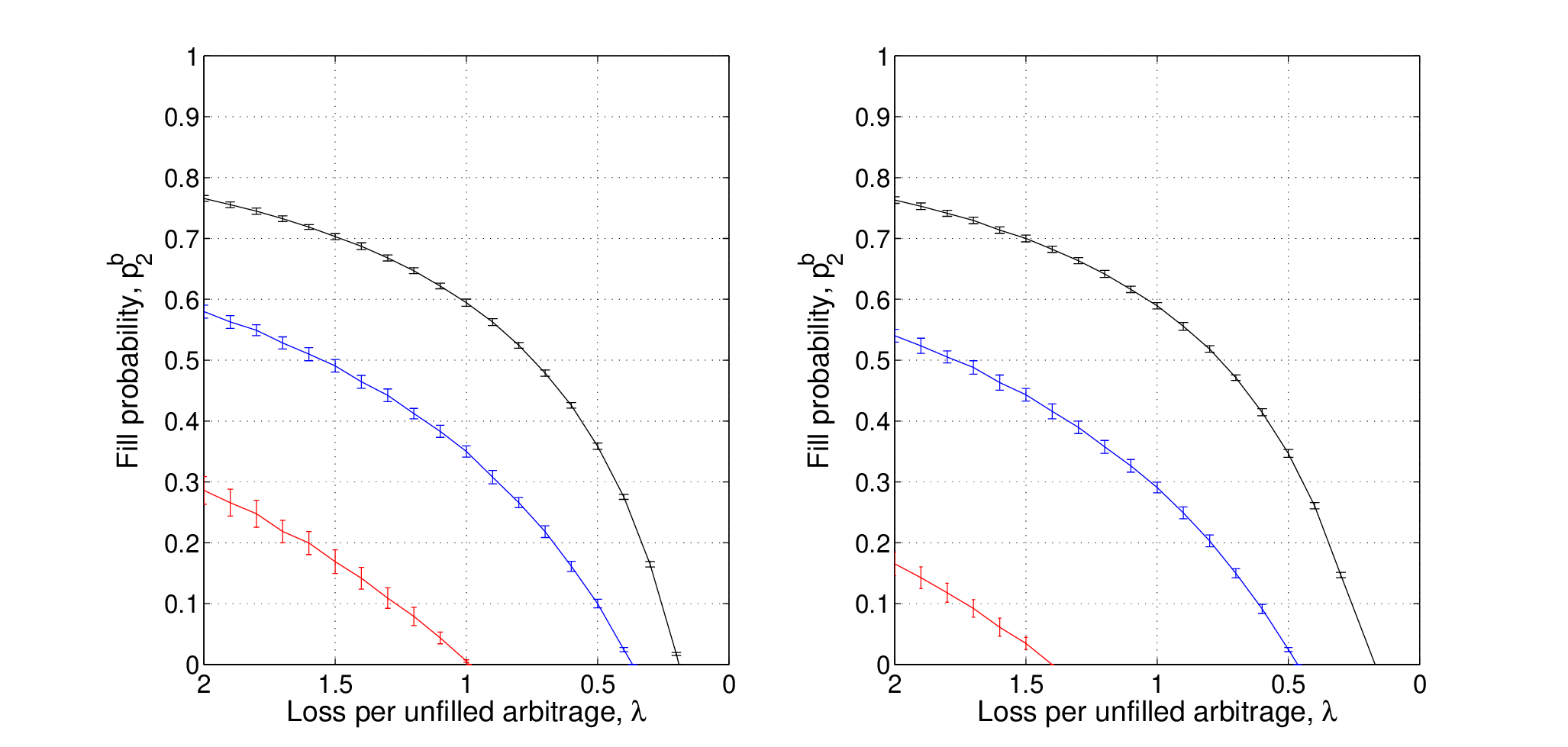,width=12.5cm}}
\vspace*{8pt}
\caption{The fill probability required to break-even as a function of the loss incurred per incomplete arbitrage transaction. Upper: scenario (1). Lower: scenario (2). Error bars indicate the standard deviation in the fill probability over $100$ simulations.}
\label{breakeven}
\end{figure}

We now investigate the fill probabilities in more detail. For scenario (1), consider a strategy trading a volume $V$ on each of $N$ arbitrage opportunities exceeding a threshold $\gamma_t$ over some time interval $W$. The total potential profit $T_1$ over this interval is then given by
\begin{eqnarray}
T_1=NV\bigg(p_1\langle\gamma-1\vert{\gamma>\gamma_t}\rangle-(1-p_1)\lambda\bigg),
\label{TP1}
\end{eqnarray}
where $\langle\cdot\rangle$ indicates an average over $W$, and the break-even fill probability $p_1^b$ (found when $T_1=0$) is given by
\begin{eqnarray}
p_1^b=\bigg(1+\dfrac{\langle\gamma-1\vert{\gamma>\gamma_t}\rangle}{\lambda}\bigg)^{-1}.
\label{bep1}
\end{eqnarray}
The break-even fill probability $p_1^b$ is therefore independent of the number of arbitrage opportunities and decreases with increasing $\langle\gamma-1\vert{\gamma>\gamma_t}\rangle$. This can be seen in Fig. \ref{breakeven} where the break-even fill probabilities are smaller for larger $\gamma_t$. For scenario (2), we take $N=n_g+n$, where $n_g$ is the number of opportunities over $W$ that last for $\ell\geq1$ second, and $n$ the number with $\ell<1$ second. The total profit $T_2$ is then given by
\begin{eqnarray}
T_2=n_gV\langle\gamma-1\vert{\gamma>\gamma_t,\ell\geq1}\rangle+nV\bigg(p\langle\gamma-1\vert{\gamma>\gamma_t,\ell<1}\rangle-(1-p_2)\lambda\bigg),
\label{TP2}
\end{eqnarray}
and the break-even fill probability by 
\begin{eqnarray}
p_2^b=\bigg(1-\dfrac{n_g\langle\gamma-1\vert{\gamma>\gamma_t,\ell\geq1}\rangle}{n\lambda}\bigg)\bigg(1+\dfrac{\langle\gamma-1\vert{\gamma>\gamma_t,\ell<1}\rangle}{\lambda}\bigg)^{-1}.
\label{bep2}
\end{eqnarray}
For this scenario, the break-even fill probability $p_2^b$ therefore depends on the proportion of arbitrage opportunities with length $\ell\geq1$, the mean value of the rate product for opportunities with length $\ell\geq1$, and the mean rate product for opportunities with $\ell<1$.

Figure \ref{breakeven} shows break-even fill probabilities generated by trading simulations and highlights the fact that $p_2^b$ is lower than  $p_1^b$, for the corresponding loss, and that the break-even fill probabilities tend to be slightly lower for the CHF than for the JPY transactions. This difference is most marked for scenario (2), with $\gamma_t=1.0001$. In this case, if a fixed loss of $2$ basis points per unfilled arbitrage is assumed, a fill probability of only $17$\% is needed to break-even.

Although this fill probability seems low, it would nevertheless be difficult to achieve. Consider a strategy where a similar fill probability of $20$\% is required to break-even. This implies that one would need to be filled on $1$ in $5$ of the arbitrage opportunities traded on. If there are $5$ market participants trading on each opportunity, each able to transact at the same speed, then this fill frequency is feasible. In the FX market, however, there are many more market participants than this competing for each arbitrage opportunity and so to achieve this fill probability one would need to identify and execute each arbitrage opportunity faster than most of these competitors. These competitors are also likely to be continually striving to increase their execution speeds in this electronic trading ``arms race''. Given the costs associated with staying ahead in this race, it would be extremely costly to maintain the fastest execution speeds and thus to regularly beat the majority of other competitors to the arbitrage prices over a prolonged period of time. The fill probabilities required to realize the profits indicated in Fig. \ref{profits} are therefore very difficult to achieve and, as a consequence, the profit levels are also extremely unrealistic.

The calculated fill probabilities also represent lower bounds of acceptability because, to justify trading on an opportunity, a trader would expect a reasonably high expected profit and not simply to break-even. When one factors in costs such as brokerage, the network connectivity required to access the market and the cost of developing and supporting a sophisticated electronic trading system, the actual fill probabilities necessary to achieve an acceptable level of profit would be substantially higher than those calculated. It therefore appears that, although mis-pricings do appear in the FX market, an unfeasibly large fill probability would need to be achieved over a prolonged period of time to realize any significant profits from them.

\section{Conclusions}
\label{conclusions}
We have shown that triangular arbitrage opportunities exist in the foreign exchange market, but that the vast majority of these opportunities are less than $1$ second in duration and $1$ basis point in magnitude. The longer, larger opportunities that do occur appear with a significantly lower frequency. We showed that, somewhat counter-intuitively, more arbitrage opportunities occur during periods of higher liquidity, but that these opportunities tend to be removed from the market very rapidly. The increased number of opportunities during liquid periods was attributed to the higher trading frequency, which resulted in more mis-pricings, but also ensured that they were quickly corrected. We have also shown that from 2003 to 2005 the market became increasingly efficient at eliminating mis-pricings and explained this by the increased use of electronic trading platforms, which enabled traders to react faster to price changes.

Finally, we used trading simulations to investigate the profitability of trading on triangular arbitrage signals. Considering the strong competition for each arbitrage, the costs of trading, and the costs required to maintain a technological advantage, it seems that a trader would need to beat other market participants to an unfeasibly large proportion of arbitrage opportunities for triangular arbitrage to remain profitable in the long-term. We therefore conclude that the foreign exchange market appears internally self-consistent. These results provide a limited verification of foreign exchange market efficiency.

\section*{Acknowledgements}
We would like to thank Mark Austin and Johannes Stolte for helpful discussions.

\end{document}